\documentclass[11pt]{article} 

\usepackage[utf8]{inputenc} 
\usepackage{cite}

\usepackage{geometry} 
\usepackage{graphicx} 
\usepackage{color}
\usepackage{soul}

\usepackage{booktabs} 
\usepackage{array} 
\usepackage{paralist} 
\usepackage{verbatim} 
\usepackage{subfig} 

\usepackage{fancyhdr} 
\usepackage{sectsty}
\allsectionsfont{\sffamily\mdseries\upshape} 

\usepackage[nottoc,notlof,notlot]{tocbibind} 
\usepackage[titles,subfigure]{tocloft} 

\newcommand{\beq}{\begin{equation}}
\newcommand{\eeq}{\end{equation}}

\newcommand{\ie}{{\em i.e., }}
\newcommand{\red}{\color{black}}

\newcommand{\bds}{\begin{displaystyle}}
\newcommand{\eds}{\end{displaystyle}}

\newcommand\myeq{\mathrel{\stackrel{\makebox[0pt]{\mbox{\normalfont\tiny def}}}{=}}}

\title{Predict or classify: The deceptive role of time-locking in brain signal classification}
\author{Marco Rusconi\footnote{dr.m.rusconi@googlemail.com}, Angelo Valleriani\footnote{angelo.valleriani@mpikg.mpg.de}\\
Max Planck Institute of Colloids and Interfaces\\
Department of Theory and Bio-Systems\\
D-14424 Potsdam, Germany}
\date{} 

\begin{document}
\maketitle
\begin{abstract}
Several experimental studies claim to be able to predict the outcome of simple decisions from brain signals measured before subjects are aware of their decision. {\red Often,} these studies use multivariate pattern recognition methods with the underlying assumption that the ability to classify the brain signal is equivalent to predict the decision itself. Here we show instead that it is possible to correctly classify a signal even if it does not contain any {\red predictive} information about the decision. {\red We first define a simple stochastic model that mimics the random decision process between two equivalent alternatives, and generate a large number of independent trials that contain no choice-predictive information. The trials are first time-locked to the time point of the final event and then classified using standard machine-learning techniques. The resulting classification accuracy is above chance level long before the time point of time-locking. We then analyze the same trials using information theory. We demonstrate that the high classification accuracy is a consequence of time-locking and that its time behavior is simply related to the large relaxation time of the process. We conclude that when time-locking is a crucial step in the analysis of neural activity patterns, both the emergence and the timing of the classification accuracy are affected by structural properties of the network that generates the signal.}
\end{abstract} 
\thispagestyle{empty}

\section*{Introduction}
The subjective feeling of consciously taking free decisions is questioned in several studies \cite{Haggard1999a,Soon2008a,Bode2011a,Soon2013a,Matsuhashi2008a,Fried2011a} inspired by the seminal works of Libet  \cite{Libet1982a, Libet1983a}. These works  \cite{Haggard1999a, Soon2008a,Soon2013a, Bode2011a, Matsuhashi2008a, Fried2011a, Libet1982a, Libet1983a} have tried to relate the onset of neural activity preceding a voluntary action with the time of decision. In {\red recent versions} of these experimental works, subjects were asked to freely decide to press a right or left button  \cite{Soon2008a, Bode2011a} while their brain activity was recorded with functional magnetic resonance imaging (fMRI). A linear support vector machine (SVM), a multivariate pattern recognition method often used to classify imaging signals  \cite{Fried2011a, Haynes2006a, HerrojoRuiz2014a}, was trained to classify left or right button-press trials. The resulting classification accuracy was found to be above chance several seconds before the time point of conscious decision  \cite{Soon2008a, Bode2011a} both in the frontopolar (BA10) and in the parietal cortex. Analogous results were obtained also in the case of complex free-decisions  \cite{Soon2013a} and supported by intracranial studies  \cite{Fried2011a}. The emergence of these so called {\em choice-predictive signals}  \cite{Soon2013a} before awareness has sparked a hot debate  \cite{Haynes2011a, Haggard2008a} within the field of neuroscience, and in research fields concerned with moral responsibility and legal culpability when the decision process occurs beyond conscious control  \cite{ResponsabilityBook}.  
\begin{figure}[h]
\begin{center}
\includegraphics[scale=.45]{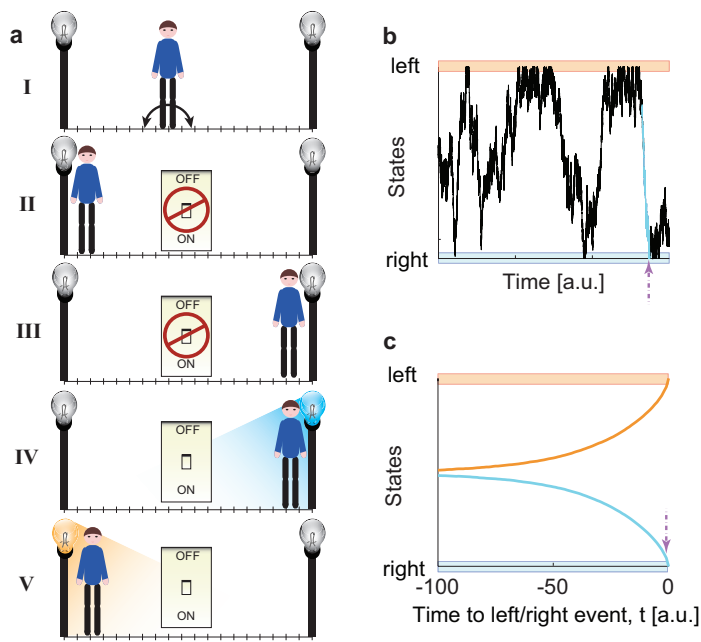} 
\end{center}
 \caption{{\bf Scheme of the model, data generation, and time-locked average.} ({\bf a}) The rules of the model. A random walker moves by discrete, equally sized steps with equal probability to the left or to the right in a room ({\bf I}). During the walk, the walker will eventually come close to the buttons placed on the right and left wall and presses the corresponding button without interrupting its walk. A switch turns on and off randomly, and independently of the position of the walker. Four types of events are possible: either the walker reaches one of the lights (left or right) when the random switch is off and the light stays off (cases {\bf II} and {\bf III}), or the switch is on and the light turns on (cases {\bf IV} and {\bf V}). ({\bf b}) The black line shows a short piece of a trajectory in which the ``right'' (blue) light on the right wall was successfully turned on. The $N$ positions, colored in light blue, preceding the successful event constitute a ``right'' trial. ({\bf c}) Averaging over many independent trials for both lights shows that the average position of the walker gets closer to the left and right walls as time grows towards $t = 0$, when the trajectories are time-locked to a successful light bulb on.}
\label{fig:1}
\end{figure}
These findings are interpreted as evidence that specific brain activity patterns contain information about the upcoming decision even before subjective awareness  \cite{Soon2008a,Bode2014a, Haynes2011a}. {\red At the base of this interpretation there is the implicit assumption that the time course of the classification accuracy is exclusively affected by information about future choices.  However, classifiers are not predictive models. To clarify the difference between classification and prediction we introduce a conceptual model that generates mutually exclusive, independent ``left" and ``right" trials to be classified first by means of SVM classification and then analyzed with information theory methods}.

The model developed here relates to the so-called WWW model, where intentional control of actions involves three main components  \cite{Brass2008b}: The ``what'' component accounts for what type of action is going to be taken, e.g., pressing the left or right button  \cite{Soon2008a, Bode2011a, Soon2013a}; the ``when'' and ``whether'' components are instead related to the timing of the action {\red and the final event of the action, respectively. }
Traditionally, the readiness potential  \cite{Kornhuber1965a} (RP) was considered to be predictive of {\it when} to move  \cite{Libet1982a, Libet1983a} and to {\red be related to the} conscious intention to move  \cite{Haggard2008a}. The origin of the RP was recently investigated using a modeling approach  \cite{Schurger2012a, Schurger2015} based on a variant of the classic drift-diffusion model  \cite{Ratcliff1978}. {\red The RP was also used in a more recent study \cite{Rusconi2015a} about vetoing voluntary decisions and shown to be a necessary but not sufficient condition for movement to take place. This is in line with those studies claiming that the RP may be a highly reproducible accident only fortuitously related to movement \cite{Schurger2015}.

In this study we introduce a model that contains the three WWW components, contains signals that are necessary but not sufficient for the final event ``left'' or ``right'' to take place, and is built in such a way that only predictions at chance level, \ie 50\%, are possible. As we shall see, after time-locking the trials to the time-point of the final event, the classification accuracy will raise well above chance level long before the end of the trials.  We will show that this happens despite the fact that the analyzed signal does not contain any choice-predictive information. }

\section*{Materials and methods}

{\red 
\subsection*{Model setup}
As a metaphor of the decision process, our model describes a random walker in a unidimensional room, \ie on a line. Time and space are discrete and the walker jumps either to the right or to the left with equal probability at each time step ({\bf Fig.\ \ref{fig:1}a}). On both opposite walls, left and right, there is one light that the walker tries to switch on by pressing one button every time he happens to reach the wall ({\bf Fig.\ \ref{fig:1}a}). The two buttons, however, are connected to power only at random times and are not synchronized either with each other or with anything else. The random process that turns the power on and off accounts for an independent {\em veto} process that aborts a decision not allowing it to become action.

The power circuit is on at each discrete time step with a fixed probability and stays on just for the duration of one time unit. If by chance the walker presses the button when the power is on, then the light will shine for just one time step. Otherwise, the light will stay off. The epoch of length $N$ of the walker's trajectory prior to a light flash corresponds to one trial ({\bf Fig.\ \ref{fig:1}b}). Here, we considered the limit case when the time between two consecutive ``power on" events is much longer than the time needed by the walker to visit the whole room uniformly. This last requirement is equivalent to ask subjects in an experiment to avoid correlations between consecutive button press events. 

To understand the effect of topology and of the intrinsic time scales, we have both considered rooms of different sizes (\ie different distances between the walls) as well as a version of the model where we give the walker the possibility to jump from any position to any other position at any step and with equal probability. As we shall see, both variants are crucial to interpret the result of the classification. An intermediate variant of the model is considered in Appendix \ref{NetwApp}. }

\subsection*{Data simulations}

{\red  In oder to quantitatively implement the metaphor discussed above and to perform the simulations of the walker in the room ({\bf Fig.\ \ref{fig:1}a}), we have considered a stationary random walk process on a line with $n+2$ positions $\{0,1,\ldots ,n,n+1\}$. One ``left'' trial is then generated as follows. {\em (i)} A very long time series of the stationary random walk is first generated in such a way that every state is visited a large number of times. {\em (ii)} Then, one out of the many occurrences of the position (or state) $0$ is chosen at random with equal probability and the $N$ preceding steps of the walk are stored.

The next ``left'' trial is generated by starting again from a new and independent time series at stationarity. The ``right'' trials are generated in a similar way by randomly choosing one of the many occurrences of the state $n+1$ instead. Each trial is extracted from an independent stationary time series that does not contain any information about which trial will be eventually extracted. In this way, the states $0$ and $n+1$ are necessary but not sufficient conditions to generate a ``left'' and ``right'' trial, respectively. Once this correspondence is determined, one can easily generalize this process to any kind of network. The easiest generalization is for the complete graph, as described below. Any other network with non-homogeneous degree distribution but symmetric with respect to $0$ and $n+1$ delivers the same qualitative results (see Appendix \ref{NetwApp}).

With this setup,} we have generated statistically independent trials simulating a random walk in discrete time on two types of networks, a linear chain  and a complete graph ({\bf Fig.\ \ref{fig:2}}). The linear chain simulates the room contained within two walls. The complete graph, instead, is a topology that guarantees the possibility for the walker to jump with equal probability to any position in just one step. {\red At steady state, the random walk visits all states $\{0,1,\ldots, n, n+1\}$, with uniform probability.} Thereby, the states $0$ and $n+1$ are two boundary states and the lights can flash only when the walker is in one of these states. {\red As already mentioned, visiting the boundary is a necessary condition for the light to go on, but it is not sufficient. We interpret this both as an effect of {\em veto} that can independently inhibit the light to shine and as a model for brain signals that are necessary but not sufficient for an event to happen.}

To mimic experimental conditions we generated $k$ independent sets (participants) of $2M$ independent trials, $M$ called ``left'' and $M$ called ``right''. Both for the linear chain and for the complete graph, we have generated the trials exploiting the time reversal property of the random walk process. {\red The same holds also for a graph with non-homogeneous degree distribution (see Appendix \ref{NetwApp}).}

\subsection*{Support-vector classification}
We used a time-resolved cross-validated linear support-vector machines (SVMs) for classification ({\bf Fig.\ \ref{ClassAppFig}}). The time point of the final event is set at $t = 0$. All other previous positions of the walker are at negative times. After time-locking at $t=0$, the $2M$ trials for each of the $k$ participants, the cross-validation approach initially consists in subdividing the trials into independent training and test groups. To avoid classification biases, training and test sets contained an equal number of ``left'' and ``right'' trials. At each time point the training set was used to train a support vector machine to distinguish between ``left'' and ``right'' decisions \cite{Soon2008a,Soon2013a, Bode2011a}. The obtained model was then used to classify the test set. We have used a leave-one-pair-out cross-validation: each pair of the $2M$ ``left'' plus ``right'' trials was successively used for testing the model learned on the remaining $2M -2$ trials. The classification performance was quantified in terms of accuracy: each cross-validation iteration produced $0\%$, $50\%$ or $100\%$ depending on whether the classifier attributed $0$, $1$ or $2$ correct labels to each pair of test trials. At each time point $t$, the goodness of classification was given as average percentage $a_t$ across all iterations ($M$ per participant). The same analysis was repeated at each time point, thus generating a time course of accuracy $a_t$ for each of the $k$ participants.  The result is presented in terms of the average across all simulations of the time course of accuracy {\red ({\bf Fig.\ \ref{fig:2}a} and {\bf \ref{fig:2}b}).} The actual classification was performed using the standard Matlab (The MathWorks, Inc., Natick, Massachusetts, United States) library for support vector machines \cite{Chang2011a}. {\red The choice of this classifier is dictated purely because we wanted to use the same analysis techniques that are commonly employed for similar experimental data, and especially those analysis techniques used in the experiment with the ``left'' and ``right'' button presses \cite{Soon2008a}.}

\subsection*{The random walks}
In this work we have considered two versions of the random walk. {\red A third version is discussed in Appendix \ref{NetwApp}}. In all cases, the state space of the walk is given by the set of $n+2$ states $\{0,1,\ldots ,n+1\}$. Technically the random walk considered here is a Markov chain in discrete time on this state space. Let $X_t $ be the random variable that gives the state visited by the process after step $t=1,2,\ldots$. The transition probabilities governing the behavior of the process are formally defined from the conditional probabilities
\beq
\label{transi}
P_{ij} \, \myeq \, \Pr\{X_{t+1}  = j \mid X_t  = i\} \, ,
\eeq
as the elements of the $(n+2)\times (n+2)$ dimensional transition probability matrix $\mathbf{P}$. We have chosen this model only because it is the simplest conceivable model conveying our main conclusions.

\subsubsection*{Random walk on a line}
The random walk on a line is the one-dimensional random walk with two boundaries. To ensure a uniform stationary distribution on this state space, the transition probability matrix $P_{ij}$ defined in (\ref{transi})  takes the following explicit form
\beq
\label{Pmatrix}
P_{ij} \, \myeq\, \left\{\begin{array}{lcl}
0.5 & {\rm if} & i=0,\, j=0,1 \\
0.5 & {\rm if} & 0<i<n+1,\, j=i\pm 1\\
0.5 & {\rm if} & i=n+1,\, j=n, n+1\, ,
\end{array}
\right.
\eeq
which can be more explicitly written as
\beq
\mathbf{P}=
\left(
\begin{array}{ccccccc}
0.5 & 0.5 & 0 & \ldots & \ldots & \ldots & 0 \\
0.5 & 0 & 0.5 & 0 &\ldots & \ldots & 0 \\
\vdots & \ddots & \ddots &\ddots & \vdots & \vdots & \vdots \\
\vdots & \vdots & \ddots &\ddots & \ddots & \vdots & \vdots \\
\vdots & \vdots & \vdots &\ddots & \ddots & \ddots & \vdots \\
0 & \ldots & \ldots & 0 & 0.5 &0 & 0.5 \\
0 & \ldots & \ldots & \ldots & 0 & 0.5 & 0.5
\end{array}\right)_{\, .}
\eeq
For later use, we define also the $t$-step transition probability matrix $\mathbf{P}^{(t)}$ defined as the $t$-th power of $\mathbf{P}$ and we indicate it with
\beq
\label{tsteps}
P_{ij}^{(t)}\, \myeq\, \Pr\{X_t =j \mid X_0 = i\}\, =\, \left(\mathbf{P}^t\right)_{ij}\, ,
\eeq
whose limit, as $t\to \infty$ gives the stationary probability distribution $\vec{\pi}$. {\red Given the choice of the transition probabilities (\ref{Pmatrix}), the stationary probability is uniform with $\pi_i=1/(n+2)$ for all $i=0,1,\ldots, n+1$.}

\subsubsection*{Random walk on a complete graph}
\label{RG}
The second random walk model considered in the manuscript is the random walk on a complete graph. On this graph, the transition probabilities are given by
\beq
\label{Pcg}
P_{ij} \, =\, \frac{1}{n+2}\, ,
\eeq
for any choice of $i$ and $j$ in the state space $\{0,1,\ldots, n+1\}$. The generation of the trials proceeds exactly as described previously.  The random walk on a complete graph has the same transition probabilities as the stationary probability distribution. Technically, the transition matrix (\ref{Pcg}) can be seen as the transition matrix (\ref{Pmatrix}) taken to an infinitely large power. {\red In fact, Eq.\ (\ref{Pcg}) is identical to the stationary distribution $\vec{\pi}$ of the process described by Eq.\ (\ref{Pmatrix}). This correspondence implies that if the time resolution of a measurement is long compared to the internal timescale, a process whose connectivity is for instance the linear chain may seem more connected than it is in reality. This correspondence does not hold for more complex networks of states with non-homogeneous degree distribution.}

\subsubsection*{Relaxation timescales}
\label{ITS}
Consider the orthogonal set of eigenvectors $\vec{v}_i$, for $i=0,1,\ldots, n+1$ of the transition matrix $\mathbf{P}$ and their associated eigenvalues $\lambda_i$, {\red whose property is that $\lambda_0=1$ and all other $\lambda_i$ have real parts strictly smaller than one in absolute value. The eigenvalue $\lambda_1$ is a real number and it is the closest to $\lambda_0$.}   

Let us define the vector of initial conditions 
\beq 
\vec{P}^{(0)}\myeq\{P^{(0)}_0, P^{(0)}_1,\ldots, P^{(0)}_{n+1}\}\, ,
\eeq
 as the vector giving the probability mass function for the stochastic variable $X_0$,
\beq
\Pr\{X_0 = j\}\, \myeq\, P^{(0)}_j\, ,
\eeq
and the vector 
\beq 
\vec{P}^{(t)}\myeq\{P^{(t)}_0, P^{(t)}_1,\ldots, P^{(t)}_{n+1}\}\, ,
\eeq
as the vector giving the probability mass function for the variable $X_t$ under the (implicit) condition that $X_0$ is distributed according to $\vec{P}^{(0)}$. Then, these two vectors are related through
\beq
\vec{P}^{(t)}\, =\, \vec{P}^{(0)}\cdot \mathbf{P}^{(t)}\, ,
\eeq
whose long time behavior gives the unique stationary probability $\vec{\pi}$ as the normalized eigenvector for the unitary eigenvalue of $\mathbf{P}$
\beq
\label{stady}
\vec{\pi}\, =\, \vec{\pi}\cdot \mathbf{P}\, ,
\eeq
where $\vec{\pi}\equiv \vec{v}_0$.
The vector $\vec{P}^{(0)}$ can be decomposed on the orthogonal space of the eigenvectors, as
\beq
\vec{P}^{(0)}\, =\, \vec{\pi}\, +\, \sum_{i=1}^{n+1} c_i \vec{v}_i\, ,
\eeq
with the $c_i$ being the projection of $\vec{P}^{(0)}$ on the vector $\vec{v}_i$. Therefore, it results that 
\beq
\label{PTvect}
\vec{P}^{(t)}\, =\, \vec{\pi}\, +\, \sum_{i=1}^{n+1} \lambda_i^t c_i \vec{v}_i\, =\, \vec{\pi}\, +\,  \sum_{i=1}^{n+1}
e^{-t/\tau_i} c_i \vec{v}_i\, ,
\eeq
where we have defined the time scales $\tau_i = -1/\log (\lambda_i)$, {\red whose real part is positive}.  Eq.\ (\ref{PTvect}) means that the vector $\vec{P}^{(t)}$ approaches the stationary state $\vec{\pi}$ as $t\to\infty$ because all $\lambda_i^t$ tend to zero in this limit. The largest of the $\lambda_i$, namely $\lambda_1$ is the one that governs the long time behavior of this limit. It is therefore customary to associate a relaxation time scale $\tau_1 $ to each Markov chain with a unique stationary state. The relaxation time scale is related to the largest eigenvalue $\lambda_1$ smaller than unity of the transition matrix $\mathbf{P}$, as 
\beq
\tau_1  \, =\,  -1/\log{\lambda_1}\, ,
\eeq
whose meaning is that $\tau_1 $ gives a lower estimate of the time scale needed to cover the state space according to the stationary probability distribution. {\red For the random walk on the line, $\lambda_1$ grows with the number of states $n$} and becomes closer and closer to the value 1, the time scale $\tau_1 $ becomes also larger and larger ({\bf Fig.\ \ref{fig:3}c}). The short time behavior of the system is however dominated by all involved time scales $\tau_i$. A consequence of this discussion is that whenever a functional depends on the elements of the $t$-step matrix $\mathbf{P}^{(t)}$ it depends on the $\lambda_i^t$ and therefore on the relaxation time scales $\tau_i$. The relaxation time scale for the complete graph is again given by the inverse of the logarithm of the largest non-trivial eigenvalue.  The relaxation time for the random walk on a complete graph is not dependent on the size of the system and is virtually zero ({\bf Fig.\ \ref{fig:3}c}). {\red This means that the process has reached the steady state just after one step, due to the fact that the one-step transition probability matrix in Eq.\ (\ref{Pcg}) is already the matrix for the stationary state. For more complex networks that interpolate between the line and the complete graph, $\lambda_1$ would depend non-trivially on both the topology of the network and on its size. The analysis of time scales presented here is very similar to the one applied in the context of neural networks \cite{Toyoizumi2015a}.}

\subsubsection*{Time reversibility}
\label{TR}
For a process at stationarity, the time reversed transition matrix is defined as
\beq
P^{(-)}_{ji}\, =\, \Pr\{X_{t-1} =i \mid X_t  = j,\, X_{0} \sim \vec{\pi}\}\, ,
\eeq
where $X_0\sim \vec{\pi}$ is a shorthand for $X_0$ being chosen according to the stationary probability mass function $\vec{\pi}$. The inverse matrix can be rearranged using the definition of conditional probabilities as
\beq
\label{inverseP}
P^{(-)}_{ji}\, =\, \frac{\pi_i}{\pi_j} P_{ij}\, ,
\eeq
where $P_{ij}$ is defined in (\ref{transi}). For both random walk models considered here, the matrix $\mathbf{P}^{(-)}$ coincides with the forward matrix $\mathbf{P}$. {\red In general, however, when the network of states has loops and cycles, the time reversibility property does not hold and the simulation of the process backward in time requires the use of Eq.\ (\ref{inverseP}).}

\subsection*{Information theory}
For a stochastic variable $Z$, which we assume here to take values in a countable set $\sigma$, we denote with $H(Z)$ its Shannon entropy, defined as
\beq
\label{Sh1}
H(Z)\, \myeq\, -\sum_{z\in \sigma} \Pr(z)\log_2 \Pr(z)\, ,
\eeq
with $\Pr(\cdot)$ being the probability mass function associated to the stochastic variable $Z$, i.e. $\Pr(z) = \Pr\{Z = z\}$ for any $z\in \sigma$. In the following we will simplify the notation and write
\beq
H(Z) \, =\, -\sum_Z \Pr(Z)\log_2 \Pr(Z)\, ,
\eeq
instead of Eq.\ (\ref{Sh1}). For the problem discussed here the mutual information
\beq
\label{Ipast}
I(S; R)\, \myeq \, H(S)\, + H(R)\, -\, H(S, R)\, ,
\eeq
asks how much information about a stimulus $S$ can be decoded from the response $R$. 

\subsubsection*{Time-locked mutual information}
Let us consider time-locked trials at $t=0$. Time-locking the trials in our random walk means that the walker's position $X_0$ at time zero is either $0$ or $n+1$ given that the light flashes. We describe this condition by saying that the response $R$ occurs at $t=0$. The time-locked mutual information $I_0(S; R)$ implements this condition through:
\beq
I_0(S; R) \myeq I(S;R\mid R \mbox{ is at } t=0)\, .
\eeq
where the response  $R$ takes values in ``left'' or ``right'' whereas $S$ is the position $X_t$ of the walker after step $t$. The calculation of $I_0$ can be reported into the framework of the standard definition given in (\ref{Ipast}) by introducing a new random variable $Y_t$ carrying the information about $X_0$ being restricted to the two boundary states $0$ and $n+1$ with equal probability ({\bf Fig.\ \ref{fig:3}a}). The variable $Y_t$ is precisely defined as the variable $X_t$ when the initial condition $X_0$ is either in state $0$ or in state $n+1$. This is more precisely expressed by the set of variables
\beq
\label{newY}
\{Y_t = k\}\, \myeq\, \{X_t=k,\, X_0=0\} \cup \{X_t=k,\, X_0=n+1\}\, ,
\eeq
which says that the event $Y_t=k$ is given by the two independent events $X_t=k$ when $X_0$ is either $0$ or $n+1$, for $k\in\{0,1,\ldots, n+1\}$, where $X_0$ is the position of the walker when the events ``left'' or ``right'' occur. Note that by construction $X_0$ must be either $0$ or $n+1$ with equal probability and that the following condition holds
\beq
\label{prop1}
\Pr\{Y_t=k\mid X_0=0\}\, =\, \Pr\{X_t=k \mid X_0=0\}\, ,
\eeq
and similar for $X_0=n+1$. Thanks to the variable $Y_t$, we have the identity
\beq
I_0(X_t;R)\, =\, I(Y_t; R)\, ,
\eeq
which allows using Eq.\ (\ref{Ipast}) for the explicit calculation.

When the time $t$ is measured relative to the time point in which the ``left'' and ``right'' trials end (both forward and backward in time), the distribution of $X_t$ is not the stationary distribution $\vec{\pi}$, derived in Eq.\ (\ref{stady}). The choice of the time-locked trials forces a new distribution of the process ({\bf Fig.\ \ref{fig:3}a)}, which leads to use an appropriate random variable defined in (\ref{newY}). We analyze now the mutual information
\beq
\label{hatI1}
I_0(X_t;R)\, = \, I(Y_t;R)\, =\, H(Y_t)\, -\, H(Y_t\mid R)\, ,
\eeq
where the subscript $0$ just reminds us that the use of the variable $Y_t$ is limited to time-locked trials, \ie on the condition $R$ occurs at time $t=0$. Here $t$ can take any integer value, with the meaning that negative $t$ means that $Y_t$ is the process before the ``left'' or ``right'' event, whereas positive $t$ means that it is afterwards. {\red When $t>0$, the time-locked mutual information is used to make predictions \cite{Palmer2015a}, when $t<0$ the time-locked mutual information can be used to perform a classification or interpolation \cite{Rudemo1975a}.} Due to the time inversion symmetry of the random walks {\red on the line and on the complete graph} there is no difference in the results for positive and for negative $t$. To increase clarity, however, we will henceforth explicit the negative sign of $t$. To proceed, we need to derive two properties of the random variable $Y$. We start with
\beq
\label{Yrule}
\begin{array}{lcl}
\bds \Pr\{Y_{-t} = k\}\eds & =& \bds \sum_{j\in\{0,n+1\}} \Pr\{X_{-t}=k\mid X_0=j\}\Pr\{X_0=j\}\eds \\
& & \\
& = &\bds \frac{1}{2}\sum_{j\in\{0,n+1\}} (\mathbf{P}^{t})_{jk}\, =\, 
\frac{1}{2}\sum_{j\in\{0,n+1\}} P^{(t)}_{jk} \eds\, ,
\end{array}
\eeq
where the transition probability matrix $\mathbf{P}$ was defined in (\ref{Pmatrix}) and where we exploit the time reversibility property {\red and Eq.\ (\ref{prop1})} and where the $t$-steps transition probabilities $P^{(t)}_{jk}$ were defined in Eq.\ (\ref{tsteps}). Given Eq.\ (\ref{Yrule}) it is now possible to define the Shannon entropy associated to $Y_{-t}$ as
\beq
\label{hatI2}
H(Y_{-t})\, =\, -\sum_{k=0}^{n+1}\Pr\{Y_{-t}=k\}\log_2\Pr\{Y_{-t}=k\}\, ,
\eeq
which is computed with elementary matrix algebra. Two limits can be easily computed by hand.  At $t=0$, the variable $Y_0$ can be either $0$ or $n+1$ with equal probability. Therefore, it results $H(Y_0)=1$. In the limit $t\to\pm\infty$, instead,  $Y_{-t}$ becomes stationary and takes the same stationary distribution $\vec{\pi}$ as $X_{-t}$ ({\bf Fig.\ \ref{fig:3}a}). In this case we obtain $H(Y_{\pm \infty})=\log_2(n+2)$. A second useful property of $Y_{-t}$ is the following
\beq
\Pr\{Y_{-t} = k\mid R = \mbox{``left''} \}\, =\, \Pr\{Y_{-t}=k\mid X_0=0\}\, =\, \Pr\{X_{-t}=k\mid X_0=0\}\, ,
\eeq
where the last probability can be computed explicitly using the transition matrix $\mathbf{P}$. A similar expression holds obviously also when $R=\mbox{``right''}$. Therefore, we can now compute the conditional Shannon entropy
\beq
\label{hatI3}
\begin{array}{rl}
\bds H(Y_{-t}\mid R) = \eds & \bds -\sum_{\alpha\in\{l,r\}}\Pr\{R=\alpha\} \sum_{k=0}^{n+1}
\Pr\{Y_{-t}=k\mid R=\alpha\}\log_2 \Pr\{Y_{-t}=k\mid R=\alpha\}\eds \\
 & \\
= & \bds - \frac{1}{2}\sum_{j\in\{0,\, n+1\}}\sum_{k=0}^{n+1}
P_{jk}^{(t)}\log_2 P_{jk}^{(t)} \eds
\end{array}
\eeq
using matrix algebra by means of Eq.\ (\ref{tsteps}). {\red In the first sum, ``left'' and ``right'' are denoted with $l$ and $r$, respectively.} Plugging Eqs.\ (\ref{hatI2}) and (\ref{hatI3}) together in the definition of $I_0$ given in Eq.\ (\ref{hatI1}) finally leads to a time dependent mutual information depending solely on the $t$-step transition probabilities. Since the time behavior of these probabilities depends only in the intrinsic timescales, the mutual information (\ref{hatI1}) decays, going backward in time, according to the time scales of the process {\red ({\bf Fig.\ \ref{fig:3}b})}. 

\subsubsection*{Unconstrained mutual information}
Also for the unconstrained mutual information $I(S; R)$, the response $R$ is one of the events ``left", ``right" and the stimulus $S$ is the position $X_t$ of the walker after step $t$. In the calculation of $I(S;R)$ there is no time-locking and the pattern $X_t$ is sampled without any knowledge about the future. The mutual information (\ref{Ipast}) can be rewritten in the more useful form 
\beq
I(X_t;R)\, =\, H(R) \, -\, H(R\mid X_t)\, .
\eeq
For the random walk models considered here, the single terms of this formula can be computed as follows. The Shannon entropy of the variable $R$ alone is given by
\beq
H(R)\, =\, -\sum_{\alpha \in \{l,r\}} \Pr\{R=\alpha\}\log_2 \Pr\{R=\alpha\}\,=\, 1\, ,
\eeq
since ``left'' and ``right'' (here denoted with $l$ and $r$, respectively) occur with equal probability. Furthermore, since $X_t$ is not time-locked with the event $R$, in our random walk the event $R$ is independent of $X_t$ by construction and thus it results also
\beq
\begin{array}{lcl}
\bds H(R\mid X_t)\eds & = &\bds -\sum_{j=0}^{n+1} \Pr\{X_t=j\} \sum_{\alpha \in \{l,r\}} \Pr\{R=\alpha\mid X_t = j\}\log_2 \Pr\{R=\alpha\mid X_t = j\}\eds \\
 & & \\
 & = & \bds \sum_{j=0}^{n+1} \Pr\{X_t=j\} H(R)\, =\, 1\, ,\eds
 \end{array}
\eeq
for any $t$ both positive and negative. Thus, putting all together it results
\beq
I(X_t;R)\, =\, 0
\eeq 
identically. This is in agreement with our expectation since the trials have been built to lead to ``left" or ``right" with equal probability independently of the values taken by the variable $X_t$. This calculation demonstrates why the unconstrained mutual information captures the true nature of the underlying process. {\red Deriving the unconstrained mutual information for our model is simple and can be done analytically. However, an application to real data may be challenging due to the limitation of imaging techniques. Such an application would indeed require the analysis of relatively large sets of data. This is necessary in order to determine the structure of the network that reproduces the dynamics connecting the various recorded patterns. For instance, fMRI measurements deliver time series of spatial brain activity patterns. After associating each spatial pattern to a state, the time series can be seen as a walk on this network of states. Once this network is known and the associated transition probabilities and the order of the Markov process describing the dynamics \cite{Bettenbuhl20122} are determined, the unconstrained mutual information can be computed.  When $X_t$ visits those states that are sufficient to generate/predict the response $R$, then the mutual information will be larger than zero. The main challenge of this method relies on the limitations of brain imaging techniques. Probably, fMRI is not suitable for this analysis since long time series must be collected, both in the presence and in the absence of the event one wants to study. However, EEG, intracranial EEG, and single cell recordings are well established methods and could allow this approach.}

\section*{{Results}}

To mimic the experimental procedure \cite{Soon2008a}, we have generated a large number of independent trials ending randomly with the event ``left" or ``right" with 50\% probability ({\bf Fig.\ \ref{fig:1}a} and {\bf \ref{fig:1}b}). The trials were built in such a way that no prediction better than 50\% is possible. Once all the trials have been time-locked at the time point of ``left" or ``right" event, they were classified using the same tools and approaches as in the experimental works \cite{Soon2008a,Soon2013a, Bode2011a} ({\bf Fig.\ \ref{ClassAppFig}}).

\begin{figure*}
\begin{center}
\includegraphics[scale=.8]{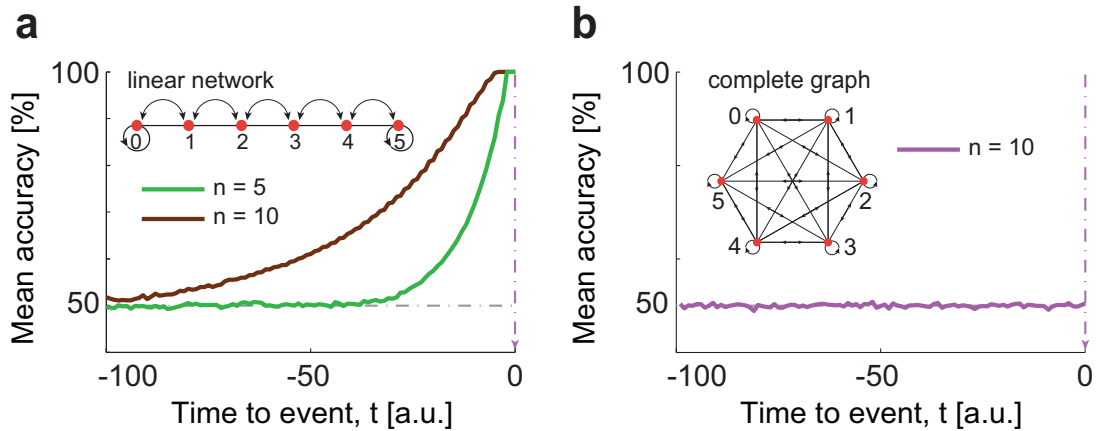} 
\end{center}
 \caption{{\bf Classification.} ({\bf a}) The solid lines show the time course of the cross-validated classification accuracy time-locked to $t = 0$ based on the average across $k=500$ realizations (\ie participants) for a random walk on a line with $n = 5$ and $n=10$. For both $n$, the accuracy decreases from 100\% moving backward from $t = 0$ and the smaller the system size is, the faster is the accuracy decrease. The inset is a scheme of the linear network. To ensure a uniform distribution, the walker can jump with equal probability either left or right from each position but the boundaries. ({\bf b}) Time course of classification accuracy for a random walk on a complete graph. The average accuracy remains 50\% and is independent of $n$. The inset shows the scheme of a complete graph, here the walker can jump from any position (or state) to any other position in just one step.}
\label{fig:2}
\end{figure*}

For trials generated using a random walk on the linear chain ({\bf Fig.\ \ref{fig:1}a}) we obtained a classification accuracy above 50\% several time steps before the end of the trials, which climbed to 100\% at $t=0$ ({\bf Fig.\ \ref{fig:2}a}). When the trials were generated using a random walk on a complete graph, instead, the mean accuracy remained at 50\% level at all times ({\bf Fig.\ \ref{fig:2}b}). If we did not know the properties of the model that generates the trials, we would have interpreted the result for the random walk on the line ({\bf Fig.\ \ref{fig:2}a}) as evidence of an activity predicting the upcoming decision while approaching the ``left" or ``right" choice. However, for both networks only predictions at chance level are possible by construction. 
Therefore, the interpretation of accuracy above chance as reflecting ``choice-predictive signals'' must be wrong and the accuracy time course calls for a different explanation. 

\begin{figure*}
\begin{center}
\includegraphics[scale=0.7]{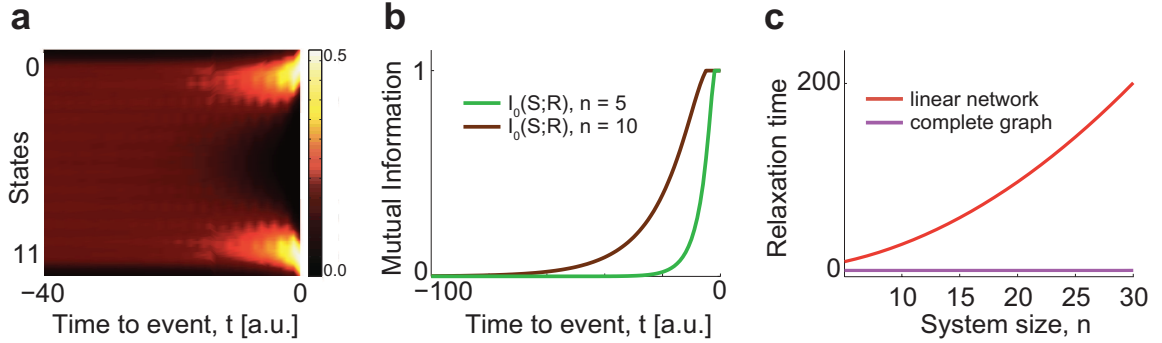} 
\end{center}
 \caption{{\bf Distribution of walker's position, mutual information and relaxation time.} ({\bf a}) The distribution of the walker's position changes from uniform to bimodal while the walker approaches the end point of the trials. At $t = 0$ the walker necessarily visits either $0$ or $(n + 1)$. The rate at which the distribution becomes uniform going backward in time from the time point $t=0$ is related to the relaxation time of the stochastic process ({\em Materials and methods}). ({\bf b}) The time-locked mutual information $I_0$ decreases moving backward in time from $t = 0$. The decrease is faster for a linear chain of smaller size because of the shorter relaxation time. ({\bf c}) For a random walk on a line the relaxation time increases with the number of states $n$. For a random walk on a complete graph the relaxation time is virtually zero and independent of the number of states. Nevertheless, the procedure to generate the ``left'' and ``right'' trials is the same for both.}
\label{fig:3}
\end{figure*}

Time-locking the trials such that the events ``left" or ``right" occur at time point $t=0$ is equivalent to knowing that at time point $t=0$ the position $X_0$ of the random walk is either equal $0$ or $n+1$ ({\bf Fig.\ \ref{fig:3}a}).
To understand the role of time-locking, we exploit decoding methods from information theory \cite{Thomas2006a, MacKay2003a}, which are intrinsically related to classification \cite{Quiroga2009a} but allow analytical treatment ({\em Materials and methods}). Methods based on information theory are often exploited to extract predictive informations from neural signals {\red especially when past stimuli are used to predict future events \cite{Palmer2015a}}. The mutual information 
\beq
I(S; R)\, \myeq\, H(S)\, + H(R)\, -\, H(S, R)\, ,
\eeq
tells how much information about a stimulus $S$ can be decoded from the response $R$, when $H(X)$ is the Shannon entropy associated to the random variable $X$ ({\em Materials and methods}). In our model, the stimulus $S$ is the position $X_t$ of the walker at time $t$ prior to the left/right event. The response $R$ is either ``left" or ``right". For times $t<0$, the time-locked mutual information 
\beq
I_0(X_t; R) \, \myeq\, I(X_t; R \mid R \mbox{ is at } t=0)\, ,
\eeq
contains the information that a response $R$ has occurred at time $t=0$.  There is a profound difference between the time-locked mutual information and the unconstrained mutual information $I = I(X_t; R)$, where no information beyond time point $t$ is known. Using methods for future conditioned stochastic processes \cite{Valleriani2008a, Li2014a, Valleriani2015a}, both functions $I$ and $I_0$ can be computed analytically for our random walk ({\em Materials and methods}). The time course of the time-locked mutual information $I_0$ ({\bf Fig.\ \ref{fig:3}b}) is qualitatively similar to the SVM classification accuracy ({\bf Fig.\ \ref{fig:2}a}): it is maximal at the time point of time-locking, \ie $t=0$, and decreases at times $t<0$ at a rate that depends on the relaxation timescales of the process ({\em Materials and methods}). In contrast to this, the unconstrained mutual information $I(X_t; R)$ is zero at all times, consistent with the fact that the random walk trajectory does not contain information about whether $R$ will be ``left" or ``right". Therefore, only the unconstrained mutual information $I(X_t; R)$ gives a faithful representation of the procedure employed to generate the trials. This result shows that time-locking combined with the slow relaxation time of the walk ({\bf Fig.\ \ref{fig:3}c}) produces classification accuracies significantly larger than 50\% before $t=0$.

 \section*{Discussion}
 
{\red Our modeling approach allowed us to understand the effect that time-locking has on the analysis of the neural signal preceding the outcome of a decision.  We have generated data with a simple strategic model and analyzed them using the standard analysis techniques, based on the SVM classifier, typically exploited in the experimental works. We have complemented the analysis with an original approach based on information theory, which allows a transparent mathematical treatment. While the accuracy of the SVM alone can be confusing, the treatment with mutual information offers more clarity and allows to highlight the conditioning introduced by time-locking. In this way, no confusion can arise. However, when one believes to compute unconstrained quantities and has overseen the conditioning introduced by time-locking, a confusion in the interpretation of the result necessarily arises. Indeed, one would erroneously come to the conclusion that the time course of the accuracy is evidence of predictive signals where instead it is just time-locking and relaxation time. We have seen, indeed, that the classification accuracy is well above the chance level of 50\% long time before the end of the trials when the trials are generated with the linear network model. We have demonstrated that this time behavior can be explained through the combined effect of network topology and relaxation timescale of the modeled process. By construction, our model does not contain any predictive information. From this we have to conclude that the raise of the classification accuracy long time before the time-locking event is not necessarily a signature of the emergence of predictive signals.

To fully capture the deceptive role of time-looking just consider the following instructive argument. Given a linear network with the buttons always connected to power, the light goes on each time one presses the button ({\bf Fig.\ \ref{fig:1}a}). If the walker is just one step before, say, the left wall, the probability that the left light will shine at the next step is $0.5$. However, if we know that the next time step will be {\it a decision time}, the same probability is $1$. This effect is reflected on the analysis and it is quantitatively evident when looking at the difference between the results of the conditioned, time-locked mutual information and the unconstrained mutual information. Only this last approach is able to show that there is no predictive signal. Our model generates a signal that is necessary but not sufficient to the generation of the final event. It may be argued that brain activity does not have such kind of signals. However, a recent experimental study on vetoing \cite{Rusconi2015a} has shown that there are necessary but not sufficient brain activity patterns related to the decision and execution of simple tasks. These signals, indeed, can deceive a classifier trained to recognize brain activities related to movement.

Beyond the technical aspects, our model belongs to a broad class of models often used to study neural activity related to decision processes \cite{Schurger2012a}. In line with these models, we believe that our result has a relevance in relation to the common paradigms in the field, as we will explain here.}

\subsection*{When, What, Whether} The neural decision of ``when'' to move was recently investigated by modeling electrophysiological signals with a leaky stochastic accumulator model  \cite{Schurger2012a}, which may look somewhat similar to our model. However, our conceptual model is different. We aimed at introducing a conceptual model that captures all the fundamental ingredients of volition  \cite{Brass2008b}. We were therefore interested in accounting not only for the ``when'', but also for the ``what'' and, most importantly, for the ``whether'' decisions. For this, our model includes a stochastic process implementing the decision between ``left'' and ``right''. This process has an intrinsic dynamics and a corresponding time-scale. Moreover, the model describes the {\em veto} process represented by the stochastic switch, which does not allow to systematically translate intention into action. This approach allowed us to show the theoretical pitfalls in the debate about free-decisions. In the present work, we describe the decision process with a simple diffusion without a drift term. This point is crucial to show that time-locking introduces a bias that generates apparent predictive signals. While our model does not allow predictions better than chance, in the stochastic accumulator model \cite{Schurger2012a} the presence of a drift term ensures by construction that eventually a decision will be taken. This is equivalent to say that the information about the decision accumulates in time and therefore predictions are intrinsically possible in the drift-diffusion model. 

\subsection*{Veto process} 
 
Our model includes a {\em veto} process represented by the stochastic switch, which does not allow to systematically translate intention into action. Similarly to other studies concerned with volition \cite{Rusconi2015a}, here we use the term {\it veto} because it was traditionally introduced by Libet. However, we do not share the dualistic flavor of Libet's interpretation of this process. In contrast to Libet, who considered {\em veto} as the control of the conscious mind uncorrelated with brain activity, we believe that the {\em veto} is implemented in specific brain networks  \cite{Brass2007a,Filevich2013}. As for the stochastic accumulator model \cite{Schurger2012a}, also our conceptual model aims at describing the decision process in its pre-motor phase while {\em veto} comes at a later stage and can inhibit the motor output of decision. We considered the decision process and {\em veto} as being statistically independent. From the experiments we know that proactive inhibition can slow down motor execution  \cite{Aron2011a}. Because our model does not account for a motor-phase extended in time, we introduced {\em veto} as a binary process that can only allow or stop the execution of the intended action instead of acting as a slowing-down mechanism.

{\red \subsection*{Relationship to Libet-like experiments} } Our result supports an alternative approach to investigate the neural determinants of free-decisions. Besides confirming the bias of time-locking and suggesting a more appropriate analysis, our approach evidences the limitation of {\it Libet-like} experimental paradigms  \cite{Haggard1999a, Soon2008a,Soon2013a, Bode2011a, Matsuhashi2008a, Fried2011a, Libet1982a, Libet1983a}. Already in his original work  \cite{Libet1983a}, Libet reported that sometimes participants consciously felt the urge to move but they inhibited their action before a movement occurred. Moreover, it was recently shown that even when their decisions are predicted in real-time using brain signals preceding their actions, participants can veto their action before movement onset  \cite{Rusconi2015a}. These experimental evidences confirm that {\em veto} implicitly plays a crucial role in Libet-like experiments. From the analysis point of view, time-locking to the button press is equivalent to ignore the {\em veto} because only trials corresponding to not-vetoed actions are considered. We have shown that this approach leads to misleading results. From the experimental point of view, paradigms that simultaneously include all decisions ({\it when}, {\it what}, and {\it whether}) can make explicit the effect of {\em veto} and are therefore more ecologically valid. When analyzing the data from such paradigms, or in order to interpret previous results  \cite{Haggard1999a, Soon2008a,Soon2013a, Bode2011a, Fried2011a, Libet1982a, Libet1983a}, it is therefore fundamental, on one hand, to quantify how the different WWW components modulates each other and, on the other hand, to quantify how this modulation changes in time. 

\subsection*{Structural and topological effects} Finally, different brain regions are characterized by different intrinsic time scales probably related to the structure of the underlying neural circuit \cite{Murray2014a}. Furthermore, several experiments \cite{Soon2008a, Bode2011a, Soon2013a} show that there are brain areas in which the accuracy of classification increases very late or it does not increase at all and that some of the {\red brain regions} showing significantly large classification accuracy are also particularly large in size \cite{Haynes2011a}. Our approach allows an interpretation of these findings. The random walk teaches us that the classification accuracy can be enhanced by increasing the relaxation time of the process. The random walk on a line, e.g., has a long relaxation time that grows with the number of states ({\bf Fig.\ \ref{fig:3}c}). This type of walk generates trials that are easy to classify ({\bf Fig.\ \ref{fig:2}a}). In contract to this, trials generated from a random walk on a complete graph cannot be classified because the relaxation time is very short ({\bf Fig.\ \ref{fig:2}b}). In this latter case the accuracy remains always around chance. Thus, small, fast, and highly connected networks will lead to little increases in accuracy; large, slow, and sparsely connected networks will produce a stronger increase in accuracy. Therefore, the classification accuracy is a useful quantity to study structural properties of the neural circuit involved in the generation of task-related brain signals.  

\section*{Conclusions}

{\red Taken together, our analyses show that classifying trials ending with a decision does not imply extracting predictive information about the decision itself.  We have shown this by using the logic of a {\em reductio ad absurdum} proof. We have generated data that do not contain predictive information about the final event, \ie ``left'' or ``right'' button press, and analyzed them with a standard classifier  after time-locking the trials to the  time point of the final event. We have shown that the time-course of the classification accuracy prior to the final event is well above the chance level depending on the topology of the underlying network of states. Since by construction the data do not contain any information about the future outcome, the high level of the classification accuracy cannot be interpreted as prediction. We have then exploited a more transparent approach based on the mutual information and demonstrated that time-locking introduces a bias analogous to future-conditioning. This allowed us to claim that the time-course of the classification accuracy at $t\le 0$ is a consequence of the network's topology and of the time scales associated to the activity on the network.} Our result adds to those critical positions questioning the existence of predictive signals of volition before awareness \cite{Guggisberg2013a,Schurger2012a}  and their interpretation in terms of free-will \cite{Klemm2010a, Bode2014a}. Instead of proving the existence of choice-predictive signals, the time course of the {\red classification accuracy} can be interpreted as the signature of task-specific structural properties of local neural circuits generating the recoded brain activity. 

{\red Our analysis shows a limitation of ``reverse-time" event-related studies, in which a signal $S$ preceding a known event $R$ is analyzed retrospectively to the occurrence of $R$. In these cases, signals $S$ that are necessary but not sufficient to $R$ will seem to be necessary {\em and} sufficient.  Furthermore, the time scale of the decay, backward in time, of the classification accuracy is not necessarily related to the information about $R$ contained in $S$ but is due also to a structural component. We have shown how this structural component produces a large and long classification accuracy even in a model where the signal $S$ has by construction no information about $R$. Time-locking to $R$ generates therefore two important biases. On one hand the role of {\it veto} is bypassed; on the other hand, time-locking introduces a conditioning in the future that can create a long-time effect backward in time depending on the network topology}. As we have shown here, this second bias produces the emergence of {\red high classification accuracies also in the absence of predictive signals. Our result, however, does not apply to those studies \cite{Palmer2015a} in which the effect of an event $R$ on the {\em upcoming} signal $S$ is studied. 

We believe that a new analysis of the data, based on stochastic predictive models \cite{Crutchfield2011a}, could help providing the time course of the unconstrained mutual information. We have discussed how our model is similar to previously studied model \cite{Schurger2012a} but differs in several crucial aspects. Albeit simple, both these models capture the essential aspects of the decision process. More complex models of neural activity could and must be introduced in the future to better quantify the neural processes leading to decisions. However, also these models will have to cope with the result discussed here as long as time-locked trajectories are analyzed backward in time.} 

\section*{Acknowledgments}
We would like to thank S. Risse and C. Allefeld for useful comments at an early stage of this work.
\section*{Author contributions}
A.V. conceived the study. M.R. interpreted the study in terms of the www model. A.V. performed the mathematical derivations. M.R. performed the SVM analyses. M.R. and A.V. produced all figures. Both authors interpreted the results, wrote and approved the manuscript.

\newpage
\appendix
\renewcommand{\theequation}{\Alph{chapter}.\arabic{equation}}
\renewcommand{\thefigure}{\Alph{section}\arabic{figure}}
\setcounter{figure}{0}
\section{Support-vector classification}
\label{ClassApp}
\label{SchemeLOO1}
\begin{figure}[h!]
\begin{center}
\includegraphics[scale=0.7]{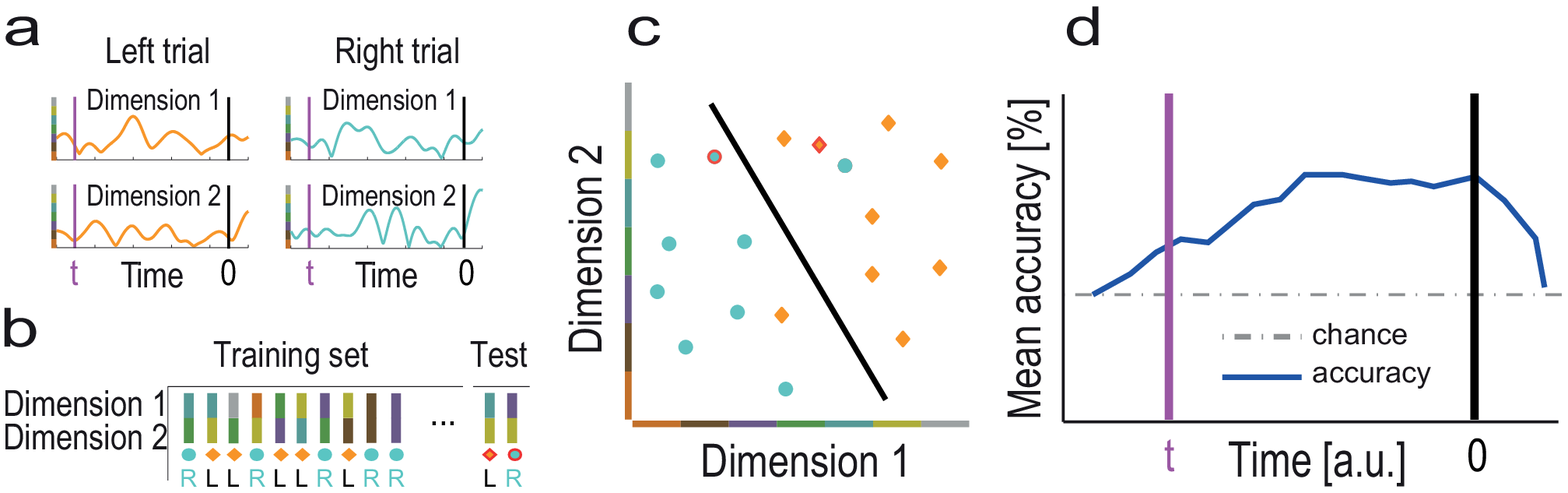} 
\end{center}
 \caption{{\bf Multivariate classification with cross validation of left/right button press trials.}
 ({\bf a}) In the first step, the trials are time-locked to the time of button press ($t=0$). The value at each time $t$ of the multidimensional imaging signal define the multivariate pattern for each of the $M+M$ trials. To exemplify, we show here one ``left" and one ``right" trial with a bivariate signal. ({\bf b}) The values at $t$ form the pattern used during the classification. All pairs of ``left" and ``right" trials but one are used to train a linear classifier to learn to distinguish the typical ``left" and ``right" trials. The remaining pair is then used to test the learned model. This procedure is repeated until each pair of trials is used for testing. ({\bf c}) The linear classifier finds a classification hyperplane using the trial set. For each of the test sets it verifies if both are on the correct side (100\% accuracy) only one is correct (50\% accuracy) or none is correct (0\% accuracy). ({\bf d}) At each $t$ the average accuracy across all $M$ tested pairs is computed. By repeating the procedure at each time point, one maps the time $t$ versus the mean accuracy with which the button press was decoded. When several participants are involved, this procedure is repeated for each participant and the outcome is then averaged. }
\label{ClassAppFig}
\end{figure}

\newpage
\section{The effect of time-locking with a complex network}
\label{NetwApp}
\setcounter{figure}{0}
In this section we discuss the effect that time-locking has on data generated with an example of a network with a more complex topology than a linear network and a complex graph. To obtain this network we introduced extra links and loops between the nodes of the linear network. The inset of Fig.\ \ref{NetwAppFig1} shows the topology that we used. After generating the data with this new network, we repeated the analysis presented in the main text. We therefore computed both the decoding accuracy, time-locked to $t=0$, and the corresponding mutual information. 

In Fig.\ \ref{NetwAppFig1} we show the time course of the decoding accuracy for the complex network and the benchmark linear network. Introducing links between the nodes allowed a more complex topology that resulted in a faster decay of the decoding accuracy while going backwards from $t=0$. The decay rate is however intermediate between linear and complete graph.     

\begin{figure}[h!]
\begin{center}
\includegraphics[scale=0.55]{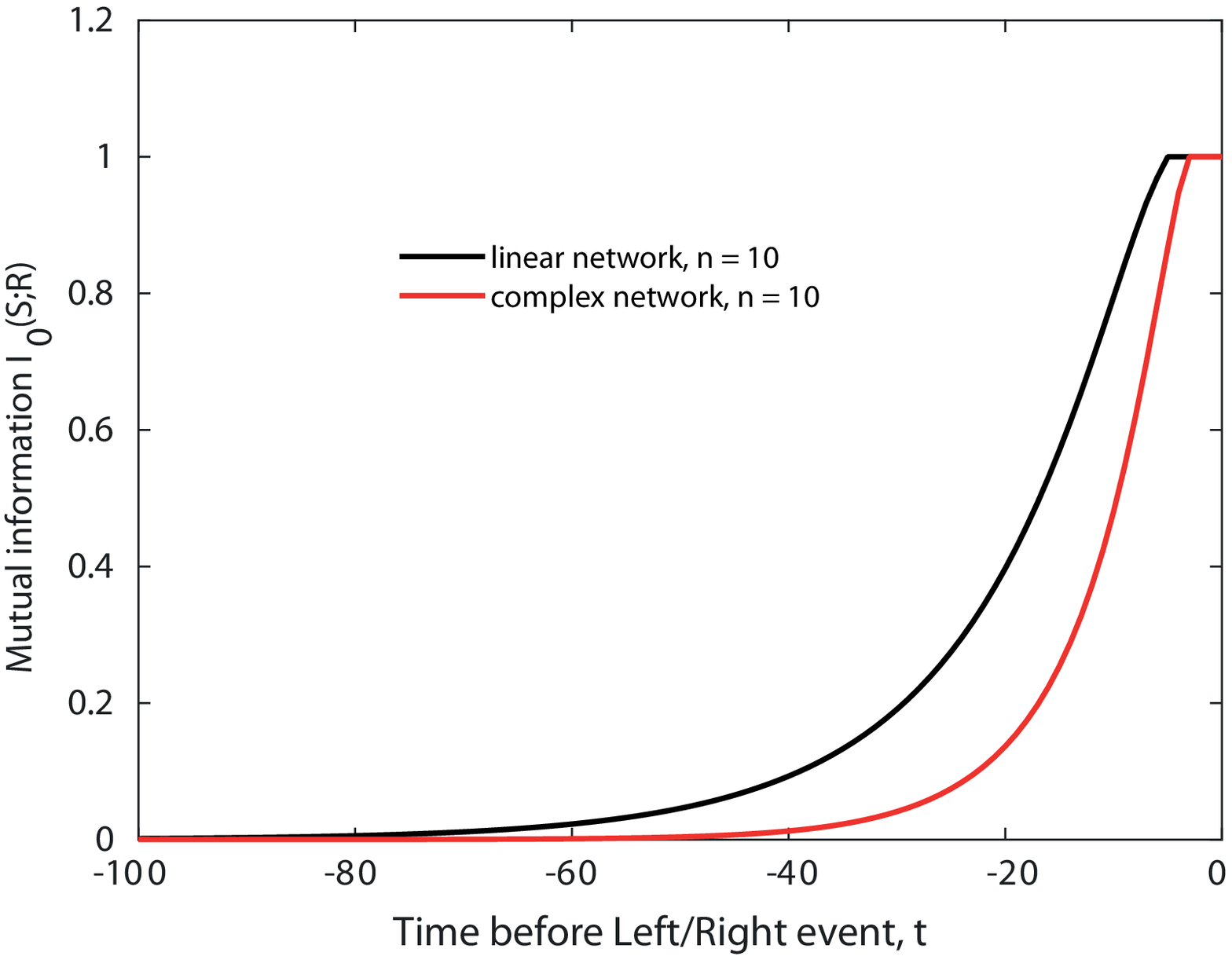} 
\end{center}
 \caption{{\bf Mutual information analysis.}
 The plot shows the time-locked mutual information $I_0(S;R)$ for the linear (black) and a complex (red) network. This last one is the same of Supplementary figure $2$. Consistently with the results reported in the main text, the decoding accuracy and the time-locked mutual information show a qualitatively similar behavior. Going backward from $t=0$ they both decay and this decays is faster for a sparse network than for a linear graph. }
\label{NetwAppFig2}
\end{figure}

In Fig.\ \ref{NetwAppFig2} we show the time course of the time-locked mutual information for the case of the complex network. Also in this case, decoding accuracy and mutual information have a qualitatively similar behavior. The unconstrained mutual information is however equal zero also in this case. This latter analysis extends to the dynamics of a complex network the results presented in the main text.

\begin{figure}[h!]
\begin{center}
\includegraphics[scale=0.55]{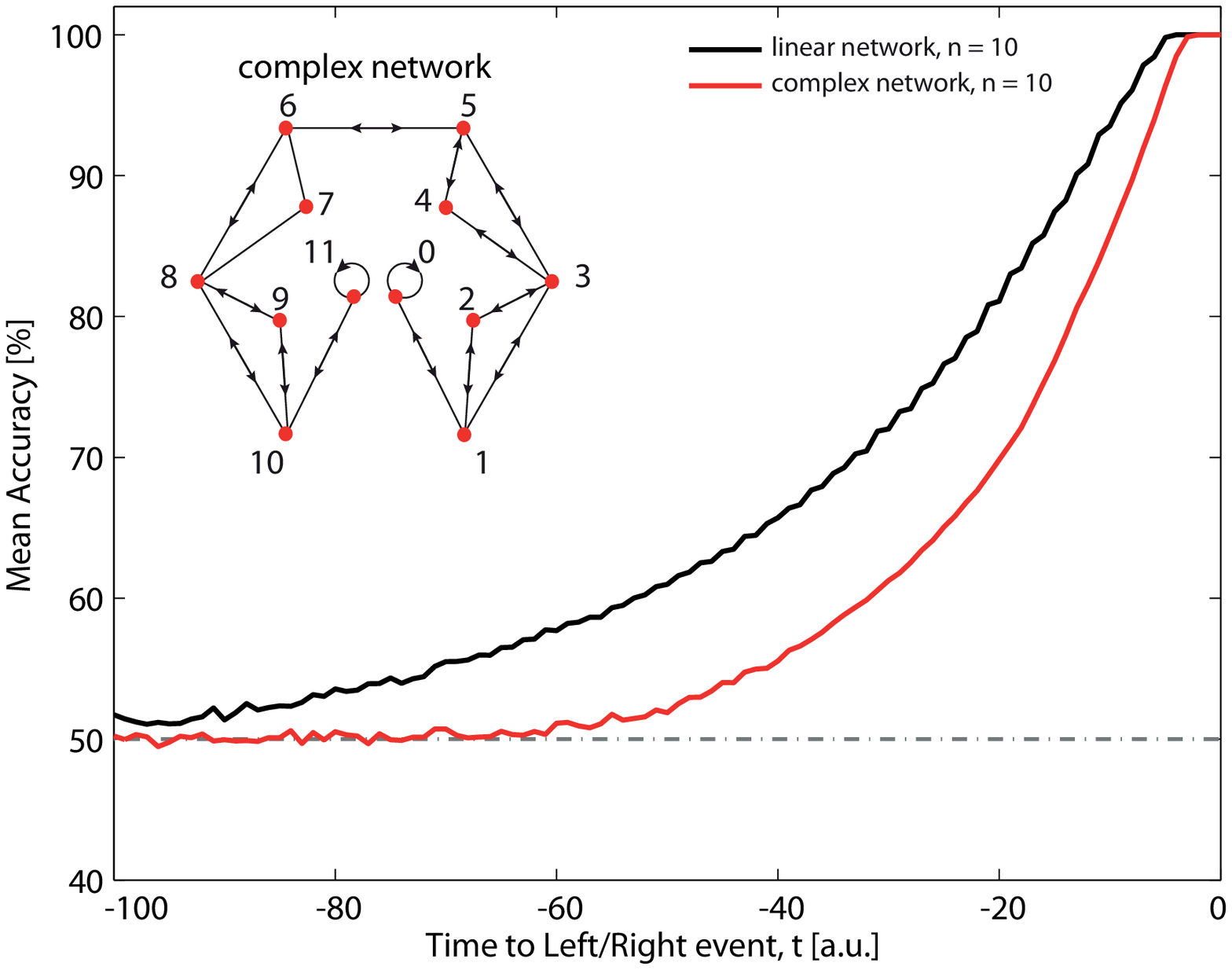} 
\end{center}
 \caption{{\bf Decoding accuracy.}
 The main figure shows the time course of the decoding accuracy for the linear (black line) and a complex network (red line) both with $n=10$ states. The inset shows the topology of the latter. Compared with the complete graph, in the complex network the topology is sparse but is more connected than the linear chain. The presence of few additional links results in a faster decaying to chance level (50\%) of the decoding accuracy moving backward from $t=0$.       
 }
\label{NetwAppFig1}
\end{figure}


\end{document}